\input harvmac
\input epsf
%\draftmode

%%%%%%%%%%%%%%%%%%%%%%%%%%%%%%%%%%%%%%%%%%%%%%%%%%%%%%%%%%%%%%%%%%%%%%%%
%%%%%%%%%%%%%%%%%%%%%%%%%%%%%%%%%%%%%%%%%%%%%%%%%%%%%%%%%%%%%%%%%%%%%%%%
%	Emil Martinec's macros
%
%
\def\nextline{\hfil\break}

\noblackbox
%

% Something to deal with sub-sub-sections

\def\unlockat{\catcode`\@=11}
\def\lockat{\catcode`\@=12}

\unlockat
% Something to deal with sub-sub-sections

\def\newsec#1{\global\advance\secno by1\message{(\the\secno. #1)}
\global\subsecno=0\global\subsubsecno=0\eqnres@t\noindent
{\bf\the\secno. #1}
\writetoca{{\secsym} {#1}}\par\nobreak\medskip\nobreak}
\global\newcount\subsecno \global\subsecno=0
\def\subsec#1{\global\advance\subsecno
by1\message{(\secsym\the\subsecno. #1)}
\ifnum\lastpenalty>9000\else\bigbreak\fi\global\subsubsecno=0
\noindent{\it\secsym\the\subsecno. #1}
\writetoca{\string\quad {\secsym\the\subsecno.} {#1}}
\par\nobreak\medskip\nobreak}
\global\newcount\subsubsecno \global\subsubsecno=0
\def\subsubsec#1{\global\advance\subsubsecno by1
\message{(\secsym\the\subsecno.\the\subsubsecno. #1)}
\ifnum\lastpenalty>9000\else\bigbreak\fi
\noindent\quad{\secsym\the\subsecno.\the\subsubsecno.}{#1}
\writetoca{\string\qquad{\secsym\the\subsecno.\the\subsubsecno.}{#1}}
\par\nobreak\medskip\nobreak}

\def\subsubseclab#1{\DefWarn#1\xdef
#1{\noexpand\hyperref{}{subsubsection}%
{\secsym\the\subsecno.\the\subsubsecno}%
{\secsym\the\subsecno.\the\subsubsecno}}%
\writedef{#1\leftbracket#1}\wrlabeL{#1=#1}}% Macros for boxes
\lockat

%

%% MORE MACROS

\def\CN {{\cal N}}

\def\CO {{\cal O}}

\def\CS {{\cal S }}

\font\manual=manfnt \def\dbend{\lower3.5pt\hbox{\manual\char127}}

\def\IZ{\relax\ifmmode\mathchoice
{\hbox{\cmss Z\kern-.4em Z}}{\hbox{\cmss Z\kern-.4em Z}}
{\lower.9pt\hbox{\cmsss Z\kern-.4em Z}}
{\lower1.2pt\hbox{\cmsss Z\kern-.4em Z}}\else{\cmss Z\kern-.4em
Z}\fi}

\def\p{\partial}

\def\CN {{\cal N}}

\def\CO {{\cal O}}
\def\CX{{\cal X}}

\def\CS {{\cal S }}

% more macros, alphabetically

\def\IZ{\relax\ifmmode\mathchoice
{\hbox{\cmss Z\kern-.4em Z}}{\hbox{\cmss Z\kern-.4em Z}}
{\lower.9pt\hbox{\cmsss Z\kern-.4em Z}}
{\lower1.2pt\hbox{\cmsss Z\kern-.4em Z}}\else{\cmss Z\kern-.4em
Z}\fi}
\def\IB{\relax{\rm I\kern-.18em B}}
\def\IC{{\relax\hbox{$\inbar\kern-.3em{\rm C}$}}}
\def\ID{\relax{\rm I\kern-.18em D}}
\def\IE{\relax{\rm I\kern-.18em E}}
\def\IF{\relax{\rm I\kern-.18em F}}
\def\IG{\relax\hbox{$\inbar\kern-.3em{\rm G}$}}
\def\IGa{\relax\hbox{${\rm I}\kern-.18em\Gamma$}}
\def\IH{\relax{\rm I\kern-.18em H}}
\def\II{\relax{\rm I\kern-.18em I}}
\def\IK{\relax{\rm I\kern-.18em K}}
\def\IP{\relax{\rm I\kern-.18em P}}
\def\IQ{\relax\hbox{$\inbar\kern-.3em{\rm Q}$}}

\def\inbar{\,\vrule height1.5ex width.4pt depth0pt}

\def\mod{{\rm mod}}
\def\p{\partial}

\font\cmss=cmss10 \font\cmsss=cmss10 at 7pt
\def\IR{\relax{\rm I\kern-.18em R}}

\def\Tr{{\rm Tr}}

% Macros for boxes
%
%\def\boxit#1{\vbox{\hrule\hbox{\vrule\kern8pt
%\vbox{\hbox{\kern8pt}\hbox{\vbox{#1}}\hbox{\k
%\hbox{$\displaystyle #1$}\kern8pt}\kern8pt\vrule}\hrule}}}
%
%%% MACROS FOR BOX BOUNDARY CONDS
%%% FROM KAWAI ET AL

\def\makeblankbox#1#2{\hbox{\lower\dp0\vbox{\hidehrule{#1}{#2}%
   \kern -#1% overlap rules
   \hbox to \wd0{\hidevrule{#1}{#2}%
      \raise\ht0\vbox to #1{}% vrule height
      \lower\dp0\vtop to #1{}% vrule depth
      \hfil\hidevrule{#2}{#1}}%
   \kern-#1\hidehrule{#2}{#1}}}%
}%
\def\hidehrule#1#2{\kern-#1\hrule height#1 depth#2 \kern-#2}%
\def\hidevrule#1#2{\kern-#1{\dimen0=#1\advance\dimen0 by #2\vrule
    width\dimen0}\kern-#2}%
\def\openbox{\ht0=1.2mm \dp0=1.2mm \wd0=2.4mm  \raise 2.75pt
\makeblankbox {.25pt} {.25pt}  }

\def\bun#1/#2{\leavevmode
   \kern.1em \raise .5ex \hbox{\the\scriptfont0 #1}%
   \kern-.1em $/$%
   \kern-.15em \lower .25ex \hbox{\the\scriptfont0 #2}%
}

\def\opensquare{\ht0=3.4mm \dp0=3.4mm \wd0=6.8mm  \raise 2.7pt
\makeblankbox {.25pt} {.25pt}  }

%%%%%%%%%%%%%%%%%%%%%%%

\def\sector#1#2{\ {\scriptstyle #1}\hskip 1mm
\mathop{\opensquare}\limits_{\lower 1mm\hbox{$\scriptstyle#2$}}\hskip 1mm}

\def\tsector#1#2{\ {\scriptstyle #1}\hskip 1mm
\mathop{\opensquare}\limits_{\lower 1mm\hbox{$\scriptstyle#2$}}^\sim\hskip 1mm}
%%%
%%%

%% ANOTHER SET OF MACROS

\def\inbar{\,\vrule height1.5ex width.4pt depth0pt}

\def\p{\partial}

\font\cmss=cmss10 \font\cmsss=cmss10 at 7pt
\def\IR{\relax{\rm I\kern-.18em R}}

%% new macros

\def\frac#1#2{{#1\over#2}}

\def\inbar{\,\vrule height1.5ex width.4pt depth0pt}
\def\IC{\relax\hbox{$\inbar\kern-.3em{\rm C}$}}
\def\IR{\relax{\rm I\kern-.18em R}}
\def\IP{\relax{\rm I\kern-.18em P}}

%
%%%%%%%%%%%%%%%%%%%%%%%%%%%%%%%%%%%%
%
\catcode`\@=11
\def\slash#1{\mathord{\mathpalette\c@ncel{#1}}}
\overfullrule=0pt

\def\CC{{\cal C}}

\def\II{{\cal I}}

\def\OO{{\cal O}}

\def\XX{{\cal X}}

\def\underrel#1\over#2{\mathrel{\mathop{\kern\z@#1}\limits_{#2}}}

\catcode`\@=12

%%%%%%%%%%%%%%%%%%%%%%%%%%%%%%%%%%%%%%%%%%%%%%%%%%%%%%%%%%%%%%

%

\def\tr{{\rm tr}}
\def\mod{{\rm mod}}

\def\exp{{\rm exp}}

%%%%%%%%%%%%%%%%%%%%%%%%%%%%%%%%%%%%%%%%%%%%%%%%%%%%%%%%%%%%%%
% new defs:

\def\vv{{\bf v}}
\def\tc{{\tilde c}}
\def\tx{{\tilde x}}
\def\ra{{\rightarrow}}
\def\HJ{{\rm Hirzebruch-Jung}}
\def\mn{{\rm mod}\; n}
%
%%%%%%%%%%%%%%%%%%%%%%%%%%%%%%%%%%%%%%%%%%%%%%%%%%%%%%%%%%%%%%%%%%%%%%%%%%

%% END MACROS
%%

%%%%%%%%%%%%%%%%%%%%%%%%%%%%%%%%%%%%%%%%%%%%%%%%%%%%%%%%%%%%%%%%%%%%%%%%%%
%%%%%%%%%%%%%%%%%%%%%%%%%%%%%%%%%%%%%%%%%%%%%%%%%%%%%%%%%%%%%%%%%%%%%%%%%%
%%%
%%% References
%%%
%%%%%%%%%%%%%%%%%%%%%%%%%%%%%%%%%%%%%%%%%%%%%%%%%%%%%%%%%%%%%%%%%%%%%%%%%%
%%%%%%%%%%%%%%%%%%%%%%%%%%%%%%%%%%%%%%%%%%%%%%%%%%%%%%%%%%%%%%%%%%%%%%%%%%

%\AdamsSV
\lref\aps{
A.~Adams, J.~Polchinski and E.~Silverstein,
``Don't panic! Closed string tachyons in ALE space-times,''
JHEP {\bf 0110}, 029 (2001)
[arXiv:hep-th/0108075].
%%CITATION = HEP-TH 0108075;%%
}

%\HarveyWM
\lref\hkmm{
J.~A.~Harvey, D.~Kutasov, E.~J.~Martinec and G.~Moore,
``Localized tachyons and RG flows,''
arXiv:hep-th/0111154.
%%CITATION = HEP-TH 0111154;%%
}

%\DixonQV
\lref\DixonQV{
L.~J.~Dixon, D.~Friedan, E.~J.~Martinec and S.~H.~Shenker,
``The Conformal Field Theory Of Orbifolds,''
Nucl.\ Phys.\ B {\bf 282}, 13 (1987).
%%CITATION = NUPHA,B282,13;%%
}

%\MartinecWG
\lref\mm{
E.~J.~Martinec and G.~Moore,
``On decay of K-theory,''
arXiv:hep-th/0212059.
%%CITATION = HEP-TH 0212059;%%
}

%\VafaRA
\lref\VafaRA{
C.~Vafa,
``Mirror symmetry and closed string tachyon condensation,''
arXiv:hep-th/0111051.
%%CITATION = HEP-TH 0111051;%%
}

\lref\mt{
%\MinwallaHJ
%\lref\MinwallaHJ{
S.~Minwalla and T.~Takayanagi,
``Evolution of D-branes under closed string tachyon condensation,''
JHEP {\bf 0309}, 011 (2003)
[arXiv:hep-th/0307248].
%%CITATION = HEP-TH 0307248;%%
}

%\DavidVM
\lref\DavidVM{
J.~R.~David, M.~Gutperle, M.~Headrick and S.~Minwalla,
``Closed string tachyon condensation on twisted circles,''
JHEP {\bf 0202}, 041 (2002)
[arXiv:hep-th/0111212].
%%CITATION = HEP-TH 0111212;%%
}

%\DouglasHQ
\lref\DouglasHQ{
M.~R.~Douglas and B.~Fiol,
``D-branes and discrete torsion. II,''
arXiv:hep-th/9903031.
%%CITATION = HEP-TH 9903031;%%
}
%\BerkoozIS
\lref\BerkoozIS{
M.~Berkooz and M.~R.~Douglas,
``Five-branes in M(atrix) theory,''
Phys.\ Lett.\ B {\bf 395}, 196 (1997)
[arXiv:hep-th/9610236].
%%CITATION = HEP-TH 9610236;%%
}
%\BerkoozKM
\lref\BerkoozKM{
M.~Berkooz, M.~R.~Douglas and R.~G.~Leigh,
``Branes intersecting at angles,''
Nucl.\ Phys.\ B {\bf 480}, 265 (1996)
[arXiv:hep-th/9606139].
%%CITATION = HEP-TH 9606139;%%
}
%\BilloYB
\lref\bcr{
M.~Billo, B.~Craps and F.~Roose,
``Orbifold boundary states from Cardy's condition,''
JHEP {\bf 0101}, 038 (2001)
[arXiv:hep-th/0011060].
%%CITATION = HEP-TH 0011060;%%
}

\lref\dm{
%\lref\DouglasSW{
M.~R.~Douglas and G.~W.~Moore,
``D-branes, Quivers, and ALE Instantons,''
arXiv:hep-th/9603167.
%%CITATION = HEP-TH 9603167;%%
}

%\HoriCK
\lref\HoriCK{
K.~Hori, A.~Iqbal and C.~Vafa,
``D-branes and mirror symmetry,''
arXiv:hep-th/0005247.
%%CITATION = HEP-TH 0005247;%%
}

%\HoriIC
\lref\HoriIC{
K.~Hori,
``Linear models of supersymmetric D-branes,''
arXiv:hep-th/0012179.
%%CITATION = HEP-TH 0012179;%%
}
%\MartinecWG
\lref\MartinecWG{
E.~J.~Martinec and G.~Moore,
``On decay of K-theory,''
arXiv:hep-th/0212059.
%%CITATION = HEP-TH 0212059;%%
}

%\MartinecTZ
\lref\MartinecTZ{
E.~J.~Martinec,
``Defects, decay, and dissipated states,''
arXiv:hep-th/0210231.
%%CITATION = HEP-TH 0210231;%%
}

%%%%%%%%%%%%%%%%%%%%%%%%%%%%%%%%%%%%%%%%%%%%%%%%%%%%%%%%%%%%%%%%%%%%%%%%%%
%%%%%%%%%%%%%%%%%%%%%%%%%%%%%%%%%%%%%%%%%%%%%%%%%%%%%%%%%%%%%%%%%%%%%%%%%%
%%% 
%%% References 
%%%
%%%%%%%%%%%%%%%%%%%%%%%%%%%%%%%%%%%%%%%%%%%%%%%%%%%%%%%%%%%%%%%%%%%%%%%%%%
%%%%%%%%%%%%%%%%%%%%%%%%%%%%%%%%%%%%%%%%%%%%%%%%%%%%%%%%%%%%%%%%%%%%%%%%%%

\lref\fulton{W. Fulton, {\it Introduction to Toric Varieties},
Annals of Mathematics Studies, vol. 131; Princeton Univ. Press (1993).} 
\lref\bpv{W.~Barth, C.~Peters, A.~Van de Ven, {\it Compact Complex Surfaces};
Springer-Verlag (1984).} 
\lref\stevens{J. Stevens, ``On the versal deformation
of cyclic quotient singularities'',
in {\it Singularity theory and its applications, part I},
LNM 1462 pp.302-319.}
\lref\ishii{A. Ishii, ``On McKay correspondence
for a finite small subgroup of GL(2,C)'',
to appear in J. Reine Ang. Math.
(available at \nextline
http://www.kusm.kyoto-u.ac.jp/preprint/preprint2000.html).}
\lref\wunram{J. Wunram, ``Reflexive modules on quotient surface
singularities'', Math. Ann. {\bf 279}, 583 (1988).}
\lref\riemenschneider{O. Riemenschneider,
``Special representations and the two-dimensional McKay correspondence''
(available at \nextline
http://www.math.uni-hamburg.de/home/riemenschneider/hokmckay.ps).}
%
%\AdamsSV
\lref\AdamsSV{
A.~Adams, J.~Polchinski and E.~Silverstein,
``Don't panic! Closed string tachyons in ALE space-times,''
JHEP {\bf 0110}, 029 (2001)
arXiv:hep-th/0108075.
%%CITATION = HEP-TH 0108075;%%
}
%
%\HarveyWM
\lref\HarveyWM{
J.~A.~Harvey, D.~Kutasov, E.~J.~Martinec and G.~Moore,
``Localized tachyons and RG flows,''
arXiv:hep-th/0111154.
%%CITATION = HEP-TH 0111154;%%
}
%
%\VafaRA
\lref\VafaRA{
C.~Vafa,
``Mirror symmetry and closed string tachyon condensation,''
arXiv:hep-th/0111051.
%%CITATION = HEP-TH 0111051;%%
}
%
%\MartinecTZ
\lref\MartinecTZ{
E.~J.~Martinec,
``Defects, decay, and dissipated states,''
arXiv:hep-th/0210231.
%%CITATION = HEP-TH 0210231;%%
}
%
%\MorrisonFR
\lref\MorrisonFR{
D.~R.~Morrison and M.~Ronen Plesser,
``Summing the instantons: Quantum cohomology 
and mirror symmetry in toric varieties,''
Nucl.\ Phys.\ B {\bf 440}, 279 (1995)
hep-th/9412236.
%%CITATION = HEP-TH 9412236;%%
}
%
%\CecottiVB
\lref\CecottiVB{
S.~Cecotti and C.~Vafa,
``Exact results for supersymmetric sigma models,''
Phys.\ Rev.\ Lett.\  {\bf 68}, 903 (1992)
hep-th/9111016.
%%CITATION = HEP-TH 9111016;%%
}
%
%\WittenYC
\lref\WittenYC{
E.~Witten,
``Phases of N = 2 theories in two dimensions,''
Nucl.\ Phys.\ B {\bf 403}, 159 (1993)
hep-th/9301042.
%%CITATION = HEP-TH 9301042;%%
}
%
%\HoriCK
\lref\HoriCK{
K.~Hori, A.~Iqbal and C.~Vafa,
``D-branes and mirror symmetry,''
arXiv:hep-th/0005247.
%%CITATION = HEP-TH 0005247;%%
}
%
%\HoriKT
\lref\HoriKT{
K.~Hori and C.~Vafa,
``Mirror symmetry,''
arXiv:hep-th/0002222.
%%CITATION = HEP-TH 0002222;%%
}
%
%\HoriFJ
\lref\HoriFJ{
K.~Hori,
``Mirror symmetry and some applications,''
arXiv:hep-th/0106043.
%%CITATION = HEP-TH 0106043;%%
}

%\HoriIC
\lref\HoriIC{
K.~Hori,
``Linear models of supersymmetric D-branes,''
arXiv:hep-th/0012179.
%%CITATION = HEP-TH 0012179;%%
}
%
%\HellermanBU
\lref\HellermanBU{
S.~Hellerman, S.~Kachru, A.~E.~Lawrence and J.~McGreevy,
``Linear sigma models for open strings,''
JHEP {\bf 0207}, 002 (2002)
arXiv:hep-th/0109069.
%%CITATION = HEP-TH 0109069;%%
}
%
%\LercheUY
\lref\LercheUY{
W.~Lerche, C.~Vafa and N.~P.~Warner,
``Chiral Rings In N=2 Superconformal Theories,''
Nucl.\ Phys.\ B {\bf 324}, 427 (1989).
%%CITATION = NUPHA,B324,427;%%
}
%
%\DouglasSW
\lref\DouglasSW{
M.~R.~Douglas and G.~W.~Moore,
``D-branes, Quivers, and ALE Instantons,''
arXiv:hep-th/9603167.
%%CITATION = HEP-TH 9603167;%%
}
%
%\HarveyNA
\lref\HarveyNA{
J.~A.~Harvey, D.~Kutasov and E.~J.~Martinec,
``On the relevance of tachyons,''
arXiv:hep-th/0003101.
%%CITATION = HEP-TH 0003101;%%
}
%
%\SenMD
\lref\SenMD{
A.~Sen,
``Supersymmetric world-volume action for non-BPS D-branes,''
JHEP {\bf 9910}, 008 (1999)
arXiv:hep-th/9909062.
%%CITATION = HEP-TH 9909062;%%
}
%
%\SenXM
\lref\SenXM{
A.~Sen,
``Universality of the tachyon potential,''
JHEP {\bf 9912}, 027 (1999)
arXiv:hep-th/9911116.
%%CITATION = HEP-TH 9911116;%%
}
%
%\KutasovQP
\lref\KutasovQP{
D.~Kutasov, M.~Marino and G.~W.~Moore,
``Some exact results on tachyon condensation in string field theory,''
JHEP {\bf 0010}, 045 (2000)
arXiv:hep-th/0009148.
%%CITATION = HEP-TH 0009148;%%
}
%
%\GerasimovZP
\lref\GerasimovZP{
A.~A.~Gerasimov and S.~L.~Shatashvili,
``On exact tachyon potential in open string field theory,''
JHEP {\bf 0010}, 034 (2000)
arXiv:hep-th/0009103.
%%CITATION = HEP-TH 0009103;%%
}

\lref\reid{
M.~Reid,
``La correspondance de McKay,''
S\'eminaire Bourbaki, 52\`eme ann\'ee, novembre 1999, no. 867, 
to appear in Ast\'erisque 2000
arXiv:alg-geom/9911165. For further references see 
http://www.maths.warwick.ac.uk/$\scriptstyle\sim$miles/McKay/
}

%\ReidZY
\lref\ReidZY{
M.~Reid,
``McKay correspondence,''
arXiv:alg-geom/9702016.
%%CITATION = ALG-GEOM 9702016;%%
}

%\MayrAS
\lref\MayrAS{
P.~Mayr,
``Phases of supersymmetric D-branes on Kaehler 
manifolds and the McKay  correspondence,''
JHEP {\bf 0101}, 018 (2001)
arXiv:hep-th/0010223.
%%CITATION = HEP-TH 0010223;%%
}

%\AnselmiSM
\lref\AnselmiSM{
D.~Anselmi, M.~Billo, P.~Fre, L.~Girardello and A.~Zaffaroni,
``Ale Manifolds And Conformal Field Theories,''
Int.\ J.\ Mod.\ Phys.\ A {\bf 9}, 3007 (1994)
arXiv:hep-th/9304135.
%%CITATION = HEP-TH 9304135;%%
}

%\BuscherQJ
\lref\BuscherQJ{
T.~H.~Buscher,
``Path Integral Derivation Of Quantum Duality In Nonlinear Sigma Models,''
Phys.\ Lett.\ B {\bf 201}, 466 (1988).
%%CITATION = PHLTA,B201,466;%%
}

%\RocekPS
\lref\RocekPS{
M.~Rocek and E.~Verlinde,
``Duality, quotients, and currents,''
Nucl.\ Phys.\ B {\bf 373}, 630 (1992)
arXiv:hep-th/9110053.
%%CITATION = HEP-TH 9110053;%%
}

%\DelaOssaXK
\lref\DelaOssaXK{
X.~De la Ossa, B.~Florea and H.~Skarke,
``D-branes on noncompact Calabi-Yau manifolds: K-theory and monodromy,''
Nucl.\ Phys.\ B {\bf 644}, 170 (2002)
arXiv:hep-th/0104254.
%%CITATION = HEP-TH 0104254;%%
}

\lref\ito{
Y.~Ito,
``Special McKay correspondence,''
arXiv:alg-geom/0111314.
}

\lref\morelli{
R. Morelli, ``K theory of a toric variety,'' Adv. in Math. {\bf 100}(1993)154
}

%\AspinwallXS
\lref\AspinwallXS{
P.~S.~Aspinwall and M.~R.~Plesser,
``D-branes, discrete torsion and the McKay correspondence,''
JHEP {\bf 0102}, 009 (2001)
arXiv:hep-th/0009042.
%%CITATION = HEP-TH 0009042;%%
}
%\DiaconescuEC
\lref\DiaconescuEC{
D.~E.~Diaconescu and M.~R.~Douglas,
``D-branes on stringy Calabi-Yau manifolds,''
arXiv:hep-th/0006224.
%%CITATION = HEP-TH 0006224;%%
}
%\DiaconescuBR
\lref\DiaconescuBR{
D.~E.~Diaconescu, M.~R.~Douglas and J.~Gomis,
``Fractional branes and wrapped branes,''
JHEP {\bf 9802}, 013 (1998)
arXiv:hep-th/9712230.
%%CITATION = HEP-TH 9712230;%%
}

%\LercheVJ
\lref\LercheVJ{
W.~Lerche, P.~Mayr and J.~Walcher,
``A new kind of McKay correspondence from non-Abelian gauge theories,''
arXiv:hep-th/0103114.
%%CITATION = HEP-TH 0103114;%%
}

%\WittenCD
\lref\WittenCD{
E.~Witten,
``D-branes and K-theory,''
JHEP {\bf 9812}, 019 (1998)
arXiv:hep-th/9810188.
%%CITATION = HEP-TH 9810188;%%
}
%\GarciaCompeanRG
\lref\GarciaCompeanRG{
H.~Garcia-Compean,
``D-branes in orbifold singularities and equivariant K-theory,''
Nucl.\ Phys.\ B {\bf 557}, 480 (1999)
arXiv:hep-th/9812226.
%%CITATION = HEP-TH 9812226;%%
}
%\DelaOssaXK
\lref\DelaOssaXK{
X.~De la Ossa, B.~Florea and H.~Skarke,
``D-branes on noncompact Calabi-Yau manifolds: K-theory and monodromy,''
Nucl.\ Phys.\ B {\bf 644}, 170 (2002)
arXiv:hep-th/0104254.
%%CITATION = HEP-TH 0104254;%%
}

%\GovindarajanVI
\lref\GovindarajanVI{
S.~Govindarajan and T.~Jayaraman,
``D-branes, exceptional sheaves and quivers on 
Calabi-Yau manifolds: From Mukai to McKay,''
Nucl.\ Phys.\ B {\bf 600}, 457 (2001)
arXiv:hep-th/0010196.
%%CITATION = HEP-TH 0010196;%%
}
%\GovindarajanEF
\lref\GovindarajanEF{
S.~Govindarajan, T.~Jayaraman and T.~Sarkar,
``On D-branes from gauged linear sigma models,''
Nucl.\ Phys.\ B {\bf 593}, 155 (2001)
arXiv:hep-th/0007075.
%%CITATION = HEP-TH 0007075;%%
}

%\HeCR
\lref\HeCR{
Y.~H.~He,
``On algebraic singularities, finite graphs and D-brane gauge theories:
A  string theoretic perspective,''
arXiv:hep-th/0209230.
%%CITATION = HEP-TH 0209230;%%
}

%\TakayanagiXT
\lref\TakayanagiXT{
T.~Takayanagi,
``Tachyon condensation on orbifolds and McKay correspondence,''
Phys.\ Lett.\ B {\bf 519}, 137 (2001)
arXiv:hep-th/0106142.
%%CITATION = HEP-TH 0106142;%%
}

%\TomasielloYM
\lref\TomasielloYM{
A.~Tomasiello,
``D-branes on Calabi-Yau manifolds and helices,''
JHEP {\bf 0102}, 008 (2001)
arXiv:hep-th/0010217.
%%CITATION = HEP-TH 0010217;%%
}

%\GovindarajanVI
\lref\GovindarajanVI{
S.~Govindarajan and T.~Jayaraman,
``D-branes, exceptional sheaves and quivers on Calabi-Yau manifolds:
{}From  Mukai to McKay,''
Nucl.\ Phys.\ B {\bf 600}, 457 (2001)
arXiv:hep-th/0010196.
%%CITATION = HEP-TH 0010196;%%
}

%\BatyrevJU
\lref\BatyrevJU{
V.~V.~Batyrev and D.~I.~Dais,
``Strong Mckay Correspondence, String Theoretic Hodge Numbers And
Mirror Symmetry,''
arXiv:alg-geom/9410001.
%%CITATION = ALG-GEOM 9410001;%%
}

%\ItoZX
\lref\ItoZX{
Y.~Ito and M.~Reid,
``The McKay correspondence for finite subgroups of SL(3,C),''
arXiv:alg-geom/9411010.
%%CITATION = ALG-GEOM 9411010;%%
}

\lref\itonak{
Y.~Ito and H.~Nakajima,
``McKay correspondence and Hilbert schemes in dimension three,''
arXiv:al-geom/9802120.
}

\lref\crawthesis{
A.~Craw,
``The McKay correspondence and representations of the McKay quiver,''
Ph.D. thesis, University of Warwick; available at
http://www.math.utah.edu/~craw.
}

%\KachruAN
\lref\KachruAN{
S.~Kachru, S.~Katz, A.~E.~Lawrence and J.~McGreevy,
``Mirror symmetry for open strings,''
Phys.\ Rev.\ D {\bf 62}, 126005 (2000)
arXiv:hep-th/0006047.
%%CITATION = HEP-TH 0006047;%%
}

%\GovindarajanEF
\lref\GovindarajanEF{
S.~Govindarajan, T.~Jayaraman and T.~Sarkar,
``On D-branes from gauged linear sigma models,''
Nucl.\ Phys.\ B {\bf 593}, 155 (2001)
arXiv:hep-th/0007075.
%%CITATION = HEP-TH 0007075;%%
}

%\HellermanCT
\lref\HellermanCT{
S.~Hellerman and J.~McGreevy,
``Linear sigma model toolshed for D-brane physics,''
JHEP {\bf 0110}, 002 (2001)
arXiv:hep-th/0104100.
%%CITATION = HEP-TH 0104100;%%
}

%\GovindarajanKR
\lref\GovindarajanKR{
S.~Govindarajan and T.~Jayaraman,
``Boundary fermions, coherent sheaves and D-branes on Calabi-Yau
manifolds,''
Nucl.\ Phys.\ B {\bf 618}, 50 (2001)
arXiv:hep-th/0104126.
%%CITATION = HEP-TH 0104126;%%
}

%\DistlerYM
\lref\DistlerYM{
J.~Distler, H.~Jockers and H.~J.~Park,
``D-brane monodromies, derived categories and boundary linear sigma
models,''
arXiv:hep-th/0206242.
%%CITATION = HEP-TH 0206242;%%
}

%\KatzGH
\lref\KatzGH{
S.~Katz and E.~Sharpe,
``D-branes, open string vertex operators, and Ext groups,''
arXiv:hep-th/0208104.
%%CITATION = HEP-TH 0208104;%%
}

%\MorrisonYH
\lref\MorrisonYH{
D.~R.~Morrison and M.~Ronen Plesser,
``Towards mirror symmetry as duality 
for two dimensional abelian gauge  theories,''
Nucl.\ Phys.\ Proc.\ Suppl.\  {\bf 46}, 177 (1996)
arXiv:hep-th/9508107.
%%CITATION = HEP-TH 9508107;%%
}

\lref\kronheimer{PB. Kronheimer and H. Nakajima, 
``Yang-Mills instantons on ALE gravitational 
instantons,'' Math. Ann. {\bf 288}(1990)263}

\lref\mooresegal{G. Moore and G. Segal, unpublished. The 
material is available in lecture notes from the April 2002 Clay 
School on Geometry and Physics, Newton Institute,  and 
http://online.kitp.ucsb.edu/online/mp01/moore1.} 

\lref\iaslectures{E. Witten, in {\it Quantum Fields and Strings: A 
Course for Mathematicians}, vol. 2, P. Deligne et. al. eds. 
Amer. Math. Soc. 1999}

\lref\kapustinlectures{A. Kapustin, Lectures at the KITP workshop 
on Mathematics and Physics, August, 2003.} 

%\SinYM
\lref\SinYM{
S.~J.~Sin,
``Comments on the fate of unstable orbifolds,''
Phys.\ Lett.\ B {\bf 578}, 215 (2004)
[arXiv:hep-th/0308028].
%%CITATION = HEP-TH 0308028;%%
}

%\LeeAR
\lref\LeeAR{
S.~g.~H.~Lee and S.~J.~H.~Sin,
``Chiral rings and GSO projection in orbifolds,''
Phys.\ Rev.\ D {\bf 69}, 026003 (2004)
[arXiv:hep-th/0308029].
%%CITATION = HEP-TH 0308029;%%
}
%\LeeSS
\lref\LeeSS{
S.~Lee and S.~J.~Sin,
``Localized tachyon condensation and G-parity conservation,''
arXiv:hep-th/0312175.
%%CITATION = HEP-TH 0312175;%%
}

%\GregoryYB
\lref\GregoryYB{
R.~Gregory and J.~A.~Harvey,
``Spacetime decay of cones at strong coupling,''
Class.\ Quant.\ Grav.\  {\bf 20}, L231 (2003)
[arXiv:hep-th/0306146].
%%CITATION = HEP-TH 0306146;%%
}

%\HeadrickYU
\lref\HeadrickYU{
M.~Headrick,
``Decay of C/Z(n): Exact supergravity solutions,''
arXiv:hep-th/0312213.
%%CITATION = HEP-TH 0312213;%%
}

%\DixonIZ
\lref\DixonIZ{
L.~J.~Dixon and J.~A.~Harvey,
``String Theories In Ten-Dimensions Without Space-Time Supersymmetry,''
Nucl.\ Phys.\ B {\bf 274}, 93 (1986).
%%CITATION = NUPHA,B274,93;%%
}

%\MaldacenaKY
\lref\MaldacenaKY{
J.~M.~Maldacena, G.~W.~Moore and N.~Seiberg,
``Geometrical interpretation of D-branes in gauged WZW models,''
JHEP {\bf 0107}, 046 (2001)
[arXiv:hep-th/0105038].
%%CITATION = HEP-TH 0105038;%%
}

%\MinwallaHJ
\lref\MinwallaHJ{
S.~Minwalla and T.~Takayanagi,
``Evolution of D-branes under closed string tachyon condensation,''
JHEP {\bf 0309}, 011 (2003)
[arXiv:hep-th/0307248].
%%CITATION = HEP-TH 0307248;%%
}

%\KapustinRC
\lref\KapustinRC{
A.~Kapustin and Y.~Li,
``D-branes in topological minimal models: The Landau-Ginzburg approach,''
arXiv:hep-th/0306001.
%%CITATION = HEP-TH 0306001;%%
}
%\KapustinGA
\lref\KapustinGA{
A.~Kapustin and Y.~Li,
``Topological correlators in Landau-Ginzburg models with boundaries,''
arXiv:hep-th/0305136.
%%CITATION = HEP-TH 0305136;%%
}
%\KapustinKT
\lref\KapustinKT{
A.~Kapustin and D.~Orlov,
``Lectures on mirror symmetry, derived categories, and D-branes,''
arXiv:math.ag/0308173.
%%CITATION = MATH-AG 0308173;%%
}

%\DouglasSW
\lref\DouglasSW{
M.~R.~Douglas and G.~W.~Moore,
``D-branes, Quivers, and ALE Instantons,''
arXiv:hep-th/9603167.
%%CITATION = HEP-TH 9603167;%%
}

\lref\segal{G.B.  Segal, ``Equivariant K-Theory,''  Publ. Math.
IHES, {\bf 34}(1968)129}

%%%%%%%%%%%%%%%%%%%%%%%%%%%%%%%%%%%%%%%%%%%%%%%%%%%%%%%%%%%%%%
% paper starts here !!!

%-------------------
% title page
%-------------------
%
\Title{\vbox{\baselineskip12pt
\hbox{hep-th/0403016}
\hbox{RUNHETC-2004-04}
}}
{\vbox{\centerline{Localized Tachyons  }
\centerline{ and the } 
\centerline{Quantum McKay Correspondence}}}
\centerline{Gregory Moore and Andrei Parnachev}
\bigskip
\centerline{{\it Department of Physics, Rutgers University}}
\centerline{\it Piscataway, NJ 08854-8019, USA}
 \vskip.1in \vskip.1in \centerline{\bf Abstract}  
\noindent
The condensation of closed string tachyons localized at 
the fixed point of a $\IC^d/\Gamma$ orbifold can be
studied in the framework of renormalization group
flow in a gauged linear sigma model.
The evolution of the Higgs branch along the 
flow describes a resolution of singularities via the
process of tachyon condensation.
The study of the fate of D-branes in this process
has lead to a notion of a ``quantum McKay correspondence.'' 
This  is a hypothetical correspondence 
between   fractional branes in an 
orbifold singularity in the ultraviolet 
with the Coulomb and Higgs branch branes in the infrared.
In this paper we present some nontrivial 
evidence for this correspondence in the case $\IC^2/\IZ_n$ by relating the 
intersection form of fractional branes to that of ``Higgs branch branes,''
the latter being branes which wrap nontrivial cycles in the resolved space.

\Date{Feb. 29, 2004}
   
%\draftmode

%%%%%%%%%%%%%%%%%%%%%%%%%%%%%%%%%%%%%%%%%%%%%%%%%%%%%%%%%%%%%%%%%%%%%%
%%%%%%%%%%%%%%%%%%%%%%%%%%%%%%%%%%%%%%%%%%%%%%%%%%%%%%%%%%%%%%%%%%%%%%

\newsec{Introduction and Summary}

In \AdamsSV\ Adams, Polchinski, and Silverstein introduced the study of localized 
tachyon condensation in closed  string theories with target space of the 
form $M \times \IC^d/\Gamma $ where $M$ is Minkowski space and the orbifold 
action by a discrete subgroup $\Gamma$ of the rotation group breaks  
 supersymmetry.  These models have turned out to provide a rich 
playground for studying both closed string tachyon condensation as well 
as the behavior of branes in the presence of closed string renormalization 
group (RG) flow. 
The present paper continues the study of D-branes in these models.
We will refine a proposal of \MartinecWG\ (reviewed below) 
for resolving a paradox related to ``missing D-brane charge.''

Although the basic techniques should generalize to any $d$, and abelian $\Gamma$,
we focus primarily on the orbifold  $\IC^2/\IZ_{n(p)}$ defined by 
the identification
\eqn\orbact{
(Z_1, Z_2) \sim (\omega Z_1, \omega^p Z_2) 
}
where $\omega = e^{2\pi i /n}$ and we assume $(n,p)=1$.  Of course, string theory contains 
spacetime fermions so we must actually lift the orbifold group $\Gamma\cong \IZ_n$ to 
a subgroup of the spin 
group $\hat \Gamma \subset SU(2) \times SU(2)$. The lift $\hat \Gamma$ 
depends on $p~ \mod\; 2n$. 
We will take $p\in (-n,n)$.
As we review in section 2 below,
because we focus on the type II string in this paper, $p$ must be odd. 
We will need to impose further  constraints on $n,p$. These will be discussed
momentarily. 
 
When $\IZ_{n(p)}$ does not preserve supersymmetry, i.e. when $\Gamma$ is not a subgroup of 
$SL(2,\IC)$, then there are localized tachyons in the orbifold. The condensation 
of these tachyons can be described in several ways. The original approach of 
\AdamsSV\ was to use D-brane probes and supergravity analysis. One can also use 
techniques suited to the study of  RG flow in $\CN=(2,2)$ 2d quantum field 
theories \refs{\hkmm,\VafaRA,\DavidVM,\MartinecTZ,\MartinecWG,\SinYM}. Some basic aspects of these techniques 
are reviewed in section 2 below. The picture which emerges is that there   is an elaborate set 
of possible flows, and the   infrared (IR) limit of the RG flow  
  is a   spacetime that blows up into a 
(partial) resolution of the singularity, producing far separated ``islands
of supersymmetric vacua.'' The most complete picture thus 
far was obtained in \MartinecWG\ where it is shown that the perturbation by a  generic 
linear combination of (primitive) chiral ring operators  flows 
(sufficiently far into the IR) to a  Higgs branch of vacua described by  
 the minimal, or Hirzebruch-Jung resolution $\CX$ of the singularity $\IC^2/\IZ_{n(p)}$. 
As explained at length in \hkmm\MartinecWG\MartinecTZ\ this complex manifold is a 
generalization of the familiar ALE resolution of $\IC^2/\IZ_{n(-1)}$ orbifolds. 
In particular the origin of $\IC^2$ is blown up into a collection of 
exceptional divisors,  $\Sigma_\alpha \cong \IC P^1$, with intersection matrix 
\eqn\hjint{
- C =  \pmatrix{-a_1 & 1 & 0 & \cdots & 0 \cr
1 & -a_2 & 1 & \cdots & 0 \cr
0 & 1 & -a_3 & \cdots & 0 \cr
\cdots & \cdots & \cdots \cdots &  & \vdots \cr
0 & 0  & 0 & \cdots   & -a_r \cr}
}
where the integers $a_\alpha\ge 2$ are the partial quotients in the continued 
fraction expansion 
\eqn\contfrac{
  {n\over p_1} =~a_1-\frac{1}{~a_2-\frac{1}{~a_3-\frac{1}{~\cdots ~1/a_r}}}
	:=[a_1, \dots, a_r] .
} 
where $p_1 = p, p>0 \;{\rm and}\; p_1=p+n, p<0$. 
For each successive string of $2$'s in the partial quotients there is a 
supersymmetric island. All of the flows discussed in \AdamsSV\hkmm\VafaRA\SinYM\  
are special cases of this prescription. 
In this paper, we restrict attention 
to the type II string. Therefore, we must take care that the generators of the chiral 
ring are not projected out by the GSO projection. In section 2 
[equations (2.15) to (2.19)], we show that the requirement that 
the generators of the chiral ring survive the GSO projection is 
equivalent to the statement that $p$ is negative and that  the partial quotients $a_\alpha$ in 
\contfrac\ are all even. This is the further restriction on $n,p$ alluded to above.

It was pointed out in \MartinecWG\ that the above description of the RG flows 
leads to a paradox concerning D-brane charge (more generally, concerning 
the category of topological D-branes). At the orbifold point the boundary 
states form a lattice, which may be identified \DouglasSW\WittenCD\GarciaCompeanRG\  with the 
equivariant K-theory lattice $K_{\IZ_n}(\IC^2) \cong \IZ^n$. However, the 
above description of the resolved space suggests that the charges of   boundary 
states should span the lattice $K^0(\CX)\cong \IZ^{r+1}$. Since 
$r':= n-1-r>0$, except in the supersymmetric case, we are left with a case of 
``missing brane charge.'' The resolution proposed in \MartinecWG\ is 
that the RG flow should be formulated in terms of the gauged linear sigma 
model (GLSM) \WittenYC\iaslectures\  and that one must account for branes in 
{\it all} the quantum vacua in the IR. In particular, one must take into account 
both the Higgs branch $\CX$ as well as the Coulomb branch of vacua. 
In the present paper we shed more light on the  relation between the
fractional branes of the orbifold conformal field theory and
the Coulomb and Higgs branch branes. 
We will perform a nontrivial check on the overall picture by studying 
the intersection form on boundary states, showing how that of the fractional 
branes ``contains'' the intersection form for Higgs branch branes.  

There is a natural basis for the D-brane boundary states of the 
orbifold $\IC^2/\IZ_{n(p)}$ which is in 1-1 correspondence with the 
unitary irreps $\rho_a$ of $\IZ_n$ \DouglasSW.  We take $\rho_a(\omega) = \omega^a$. 
We denote the corresponding boundary states by $e_a$, $a=0,\dots, n-1$. 
The regular representation $e_0+\cdots + e_{n-1}$ corresponds to 
a $D0$ brane, which may freely move away from the orbifold point, while 
$e_1, \dots, e_{n-1}$ correspond to ``fractional branes,'' which are 
pinned at the orbifold point. The main result of this paper is that one 
must make a nontrivial change of basis from $e_a$ to form boundary states 
$h_0,\dots, h_r$ and $c_1, \dots, c_{r'}$
corresponding to branes in the Higgs and Coulomb  vacua, respectively. 
As explained in \MartinecWG, and reviewed in section 3 below, 
the GLSM naturally makes a distinction 
between two sets of  irreps of $\IZ_n$. The ``special representations''
  may be  identified with a single $D2$ brane 
wrapping an exceptional divisor. 
\foot{Thanks to the sequence $0 \ra \CO(-\Sigma) \ra \CO \ra \iota_*(\CO_\Sigma) \ra 0$, 
the special representations are also in correspondence with 
boundstates of a single $D4$ brane 
wrapping $\CX$ and bound to a $D2$ brane wrapping one of the exceptional 
divisors $\Sigma$. }  In this paper we refer to the ``special representations'' 
as the ``Higgs representations,'' while the $r'$ nontrivial irreducible 
representations 
in the complement will be called ``Coulomb representations.''
Denoting Higgs representations by $\alpha= 1,\dots, r$ and Coulomb 
representations by $\nu = 1, \dots, r'$ we have 
\eqn\chgbasis{
\eqalign{
h_0 & = e_0 + \cdots + e_{n-1} \cr
h_{\alpha} & = e_\alpha + \sum_{\nu=1}^{r'} u_{\alpha}^{~\nu} e_{\nu} \cr
c_{\nu} & = e_{\nu} \cr}
}
where $u_{\alpha}^{~\nu}$ is a matrix of integers. 

The main tool we employ to arrive at \chgbasis\ is the 
intersection form on boundary states. Quite generally, if 
$a,b$ are D-brane boundary conditions then the 
Witten index in the open string channel $\CH_{ab}$ 
defines a bilinear pairing $\CB \times\CB \to \IZ$, where 
$\CB$ is the set of boundary conditions \DouglasHQ\BerkoozIS\BerkoozKM. 
(We may think of these as objects in an additive category, or 
as a lattice of boundary states. If the ground ring is 
the complex numbers then the pairing is sesquilinear. ) For example, if $a,b$ 
define $K$-theory classes on a smooth spacetime then 
$(a,b) = {\rm Ind}(\Dsl^+_{\bar a\otimes b} ) $ is given by the 
index of the chiral Dirac operator. 
\foot{Our pairing will always be symmetric or antisymmetric. 
In the literature on topological branes a different,  asymmetric, pairing 
on boundary states is sometimes employed. The difference between the 
two pairings is analogous to (and sometimes, is) the difference between 
the Dirac index and the $\bar \p$ index. } 
Since the  pairing is an index it is expected to be 
invariant under (finite) RG flow. Thus, comparison 
of the pairing of fractional branes and Higgs branch branes leads to 
a nontrivial relation between these branes. 

The  quadratic form on fractional branes 
is easily computed (see  section 4 below). Let $\rho_a,\rho_b$ be representations 
of $\Gamma$ corresponding to fractional branes $a,b$, and let 
$\pi_*$ be the projection of a (virtual) representation onto the 
trivial representation. Since $\Gamma$ is lifted to 
$\hat \Gamma$ in the spin group, the chiral spin representations $S^\pm$  
restrict to representations of $\Gamma$. The intersection form   is given by
%%%%%

\eqn\kpair{
(a,b)  = \pi_*\biggl( \bar \rho_a\otimes \rho_b \otimes (S^+ - S^-) \biggr).
}
For the case $d=2$, $\Gamma=\IZ_n$ this reduces to 
\eqn\frovlp{
I = \CS^{(p-1)/2}+\CS^{-(p-1)/2}-\CS^{(p+1)/2} - \CS^{-(p+1)/2} 
}
where $\CS$ is the $n\times n$ shift matrix.  Equation \frovlp\ is to be 
contrasted with the intersection form for Higgs branch branes,
the latter being  the geometrical intersection form \hjint.
%
%\foot{
%More precisely, it is the Dirac index pairing on the pushforward sheaves
%$\iota_*(\CO_{\Sigma_\alpha})$.} 
%
The main technical step, which is explained in detail in section 
five  below, is to find a matrix of {\it integers } $u_\alpha^{~\nu}$ so that the 
change of basis \chgbasis\ brings $I$ to the block diagonal form 
\eqn\ibd{
  F=-UIU^T =\pmatrix{
~0& ~0& ~0\cr
~0& ~C& ~0\cr
~0& ~0& ~C'}
}
where $C$ is defined in \hjint. 

As explained in \MartinecWG\ the relation of fractional branes to the 
Coulomb and Higgs branes of the GLSM  suggests  a generalization of the 
McKay correspondence which was termed the ``quantum McKay correspondence.'' 
(See \MartinecWG\ for an extensive set of 
references to the McKay correspondence in the math and physics 
literature.) The name ``special representation'' indeed originates from
mathematical investigations into the generalization 
to  non-crepant (spacetime non-supersymmetric) resolutions of the two-dimensional McKay 
correspondence \wunram\ishii\ito. In the math literature it is shown 
that there is a 1-1 correspondence between special representations 
and   the tautological line bundles on $\CX$. As explained in 
\MartinecWG\ and section 3 below, this is naturally understood in 
the context of the GLSM. The important new point, which is suggested by physics,
is that by including the Coulomb branes one has a correspondence 
more analogous to the crepant case. Moreover, physics strongly 
suggests that the correspondence can be phrased at the level of 
derived categories of sheaves, and that it should apply to 
all toric resolutions of orbifold singularities. Some mathematical 
results relevant  to this conjecture 
were announced in \kapustinlectures\ and attributed 
to Orlov, but unfortunately no details have as yet been available.

We must stress an important technical restriction on 
our result. We have only managed to find a change of basis 
bringing \frovlp\ to the form \ibd\ in the case where the 
GLSM can be embedded in the GSO-projected type II string. 
As we have mentioned, this requires $p\in (-n,n)$ to be odd
and negative and the partial quotients $a_\alpha$ to be even. We  
comment on the type 0 string in section 6. 
It is not clear to us if the restriction to the type II string 
is of fundamental importance, or if it is merely a technical 
simplification. By studying  specific examples one can show 
that an analogous change of basis in the type 0 string 
will in general be much more complicated than 
\chgbasis. 
%%%%%
It is perhaps important to note in this context that 
the type 0 branes are sources of the bulk tachyon. 
 
To summarize, this paper is organized as follows. 
In the next section we briefly review the condensation of closed string
tachyons in the twisted sectors of $\IC^2/\IZ_{n(p)}$ orbifolds and
show that the ring generators survive
the GSO projection if and only if $p$ is negative and all $a_\alpha$ are even
integers.
In section 3 we recall the gauged linear sigma model 
description of tachyon condensation and give a preview
of what happens to the orbifold fractional branes, following \MartinecWG.
In particular, we explain how the Higgs representations can
be inferred from the $\IZ_n$ action.
In section 4 we study open strings living on the orbifold
and compute the  intersection matrices of the fractional branes.
The main technical work is in 
 section 5, where  we explain how one can bring $I$  to   block-diagonal form
\ibd.
We show that this is possible precisely when the chiral ring 
generators are not projected out by the chiral GSO.
We discuss type 0 theory, and some other loose ends,  in section 6.

%%%%%%%%%%%%%%%%%%%%%%%%%%%%%%%%%%%%%%%%%%%%%%%%%%%%%%%%%%%%%%%%%%%%%
%%%%%%%%%%%%%%%%%%%%%%%%%%%%%%%%%%%%%%%%%%%%%%%%%%%%%%%%%%%%%%%%%%%%%%

\newsec{Chiral ring, singularity resolution and GSO projection}

We consider superstring theory in 9+1 dimensions.
The orbifolding by $\IZ_{n(p)}$ happens in the 67 and 89 planes,
parametrized by complex coordinates $Z^{(1)}$ and  $Z^{(2)}$.
The orbifold group is generated by
\eqn\orbaction{ w=\exp\left({2\pi i\over n}(J_{67}+p J_{89})\right) ,}
where $J_{67}$ and $J_{89}$ generate rotations in two complex planes
and $p$ is defined ${\rm mod} \, 2n $.
We will take the fundamental domain to be $p\in (-n,n)$.
The action of $w^n$ on the Ramond sector ground state is a multiplication by $(-1)^{p\pm 1}$,
depending on chirality.
When $p$ is even, this acts as $(-1)^F$ where $F$ is the spacetime fermion
number.
In type II, there is no bulk tachyon and there are closed string
fermions in the bulk, hence $p$ must be odd \refs\AdamsSV.

Let us briefly review the formulation in NSR formalism.
For a more complete discussion see  \refs\hkmm.
The complex coordinate $Z^{(i)}$ is promoted to the
worldsheet chiral superfield
\eqn\chirsup{\Phi^{(i)}=Z^{(i)}+\theta_+\psi_+^{(i)}+\theta_-\psi_-^{(i)}+\cdots\quad .}
Here pluses and minuses in the subscript correspond to
the right and left movers; we reserve bar for the target
space complex conjugation.
The fermions can be bosonised
\eqn\bosfer{\psi_+^{(i)}=e^{iH_+^{(i)}};\;\;\psi_-^{(i)}=e^{i H_-^{(i)}}\ .}
The orbifold action on the worldsheet fields is
\eqn\orbphiact{w:\;\Phi^{(1)}\to e^{2\pi i\over n}\Phi^{(1)}\qquad
                     \Phi^{(2)}\to e^{2\pi i p\over n}\Phi^{(2)} .}
The theory contains $n-1$ twisted sectors, labeled by
$s=1,2,\cdots, n-1$.
We want to construct vertex operators $X_s$ that correspond to ground
states in the twisted sectors.
A useful ingredient is an operator
\eqn\defxj{
  X^{(i)}_s=\sigma_{s/n}^{(i)}\;\exp\left[i(s/n)(H_+^{(i)}-H_-^{(i)})\right]
	\quad ;\qquad s=1,2,\cdots, n-1
}
where $\sigma_{s/n}$ is the bosonic twist $s$ operator \refs\DixonQV.
In the following we will restrict our attention to the right movers
and will drop the $+$ and $-$ subscripts.
The tachyon vertex operator can now be written as
\eqn\xj{X_s=X^{(1)}_s X^{(2)}_{n\left\{ {sp\over n} \right\}}       ,}
where $\left\{ x \right\} \equiv x-\left[ x \right]$
is the fractional part of $x$.
The generators of the chiral ring $W_\alpha, \;\;\alpha=1\ldots r$ 
form a collection of (in general) relevant operators. Turning these on 
in the action induces RG flow to  
the minimal resolution of the singularity \refs{\hkmm,\MartinecWG}.
For the $\IC^2/\IZ_{n(p)}$ orbifold such a resolution is encoded
in the continued fraction
\eqn\cfdef{{n\over p_1}=a_1-{1\over a_2-{1\over a_3- \ldots}}\; := \; [a_1,a_2,\ldots a_r]      ,}
where
\eqn\tpdef{p_1=p,\; p>0;\qquad p_1=p+n,\; p<0   ,}
defines the action of the orbifold group on the bosonic
fields; $p_1\in (0,n)$.
The generators of the chiral ring are in one-to-one correspondence
with the $\IP^1$'s of the minimal resolution, which requires $r$
$\IP^1$'s with self-intersection numbers $-a_\alpha$.
The intersection number of adjacent $\IP^1$'s is equal to one.
In other words, the intersection matrix is the negative of the generalized
Cartan matrix,
\eqn\exc{C_{\alpha\beta}=
         -\delta_{\alpha,\beta-1}+a_\alpha \delta_{\alpha,\beta}-\delta_{\alpha,\beta+1}}

It is possible to label the generators of the chiral
ring by the set of integers $\{q_i\}$ and $\{p_i\}$ which
are determined by the continued fraction $[a_1,a_2,\ldots a_r]$
via the recursion relations \MartinecWG:
\eqn\rrpq{\eqalign{
p_{j-1}/p_j=&\left[a_j,a_{j+1},\ldots,a_r\right]\cr
q_{j+1}/q_j=&\left[a_j,\ldots,a_1\right]\;,\qquad 1\le j\le r
}}
with the initial conditions
$q_0=0,\;q_1=1,\;p_{r+1}=0,\;p_r=1$.
The vectors
\eqn\defv{{1 \over n} \vv_j ={1\over n} (q_j,p_j)\; ,\qquad j=0,\ldots,r+1}
satisfy the recursion relations
\eqn\rrv{ a_j\vv_j=\vv_{j-1}+\vv_{j+1}, \qquad 1\le j\le r,}
In fact, $\vv_\alpha$, $\alpha =1, \dots, r$, 
are in one-to-one correspondence with the vectors providing the
minimal resolution of the singular fan describing $\IC^2/\IZ_{n(p)}$.
They also label the generators of the chiral ring
\eqn\genj{W_\alpha =X_{q_\alpha}^{(1)} X_{p_\alpha}^{(2)}         .}
In other words, $q_j$ and $p_j$ are the charges of $U(1)_R$
symmetries generated by the currents $\psi^{(i)} {\bar \psi}^{(i)},\;i=1,2$.
The intersection matrix \exc\ encodes the ring relations between
the generators \hkmm
\eqn\wrel{W_{j-1} W_{j+1} = W_j^{a_j}, \qquad 1\le j\le r .}

In type II there is a subtlety having to do with the
existence of chiral GSO projection.
By comparing with the spectrum of the light-cone Green-Schwarz 
formulation of the orbifold, one finds that 
the  action of $(-1)^F$ is given by \refs\hkmm
\eqn\gsoact{ H^{(1)}\ra H^{(1)}+\pi p,\qquad H^{(2)}\ra H^{(2)}-\pi .}
which implies that $p$ must be odd.
\foot{The GSO projection is also discussed in \VafaRA\LeeAR\LeeSS.} 
The action on tachyon vertex operators is
\eqn\gsoactx{X_s\ra(-1)^{\left[sp\over n\right]} X_s   .}
Only odd operators in the (-1,-1) picture survive chiral GSO projection,
which means that all $X_s$ with even $\left[sp\over n\right]$ are projected out, while
those with odd $\left[sp\over n\right]$ survive.
This means that for positive $p$ at least one generator
of the chiral ring,  $W_1=X_1$, is projected out.
In the following we will be interested in the examples where
the set of generators is left intact by the GSO projection.
A necessary condition is therefore that $p$ must be negative, which
implies $p_1=p+n$.

According to \genj, \gsoact, the action of $(-1)^F$ on the generators is
\eqn\gsoactw{W_j\ra (-1)^{B'_j} W_j, }
where
\eqn\defbj{B'_j = {1 \over n}(p_j -p q_j) = B_j-q_j      .}
Here $B_j$ are integers which satisfy the same recursion
relations as $q_j$ and $p_j$ \refs\MartinecWG\
\eqn\ab{ a_j B_j=B_{j-1}+B_{j+1}, \qquad 1\le j\le r      }
with initial conditions $B_0=1,\;B_1=0$.
{}From \rrv, \defbj\ and \ab\ it follows that $B'_j$ satisfy the
same recursion relations \ab\ with boundary conditions $B'_0=1,B'_1=-1$.
According to \gsoactw, all $W_j$ survive the GSO projection if and
only if $B'_j$ are all odd, which is equivalent to all $a_j$ being
even.
As we will see below, precisely for such orbifolds the intersection
matrix of fractional branes contains the intersection matrix of the
minimal cycles in the \HJ\ resolution.

%%%%%%%%%%%%%%%%%%%%%%%%%%%%%%%%%%%%%%%%%%%%%%%%%%%%%%%%%%%%%%%%%%%%%%%
%%%%%%%%%%%%%%%%%%%%%%%%%%%%%%%%%%%%%%%%%%%%%%%%%%%%%%%%%%%%%%%%%%%%%%

\newsec{GLSM and special representations}

In \MartinecWG\ the gauged linear sigma model was used to shed light
on the fate of fractional D-branes in the process of twisted
tachyon condensation.
The basic puzzle that was adressed in \MartinecWG\ is the following.
%%%%
The charges of fractional D-branes on the orbifold are 
described by the equivariant K-theory,
which is isomorphic to $\IZ^{n}$ as a $\IZ$-module.
On the other hand, K-theory of the \HJ\ manifold is of rank 
$r+1$. The compactly supported $K$-theory is generated
by $r$ branes wrapping a basis of two-cycles, together with the $D0$ brane. 
What happens to the other $r'=n-r-1$ branes?
The gauged linear sigma model answer \MartinecWG\  is that these extra
branes live on the Coulomb branch of the theory, while it is
the Higgs branch which describes the geometry of the  resolved
space.

%%%%
The $D2$ branes wrapping the $r$ independent cycles in $\CX$ 
correspond to $r$ ``special representations'' of the orbifold group $\IZ_n$. 
Let us now recall  the special representations of the
orbifold group and how they appear. The 
gauged linear sigma model description \MartinecWG\ involves $r$ $U(1)$ gauge
fields $A_\alpha,\;\alpha=1\ldots r$ and  $r+2$ chiral
fields $X_i,\; i=0\ldots r+1$ with charges
\eqn\xcharge{Q_{i\alpha}=\delta_{i,\alpha-1}-a_\alpha\delta_{i,\alpha}+\delta_{i,\alpha+1}   .}
There are also $r$ FI terms $\zeta_\alpha$, which are renormalized.
The geometry for a given set of  $\zeta_\alpha$'s is determined by solving
$F$ and $D$-term equations and modding out by the unbroken gauge
group.
At the UV fixed point, $\zeta\ra -\infty$, this unbroken gauge group
is precisely $\IZ_n$ and the theory is reduced to the $\IC^2/\IZ_n$
orbifold.
In the IR, $\zeta\ra +\infty$, the RG flow leads to the \HJ\ space;
the non-minimal curves are blown down \MartinecWG.
This space can be viewed as a quotient of a $U(1)^r$ bundle $S_\zeta$ over the
base space $\XX$ by the natural action of the $U(1)^r$ group element
$g=(e^{i\theta_1}\ldots e^{i\theta_r})$,
\eqn\gact{ g:X_i\ra e^{i Q_{i\alpha}\theta_\alpha} X_i     .}
The K-theory of $\XX$
\eqn\kth{ K^0(\XX)=\IZ\oplus\IZ^r  .}
is generated by $\OO$, the trivial line bunlde on $\XX$, and tautological
line bundles $R_\alpha$ corresponding to D-branes filling $\XX$ with magnetic
monopoles in different exceptional divisors.
These are the vector bundles associated to the representations
$\rho_\alpha(g)=e^{i\theta_\alpha}$,
\eqn\ralpha{R_\alpha=(S_\zeta\times\IC)/U(1)^r   ,}
where the action of $g\in U(1)^r$ is given by \gact\ and
\eqn\gact{g: \upsilon\ra e^{-i\theta_\alpha}\upsilon, \qquad\upsilon\in\IC   .}
What happens to these D-banes as the FI parameters $\zeta_\alpha\ra-\infty$, where
$\XX$ becomes a $\IC^2/\IZ_n$ orbifold?
The representations $\rho_\alpha$  should now be restricted to the
unbroken gauge group $\IZ_n$.
As explained in \MartinecWG, this unbroken gauge group consists of elements
\eqn\zn{ \left(e^{i\theta_1},\ldots,e^{i\theta_r}\right)=
          \left( \exp[2\pi i {p_1 m_1\over n}],\ldots,
                 \exp[2\pi i {p_r m_r\over n}  ] \right)   }
where $m_\alpha \in\IZ$.
Hence, $R_\alpha$'s correspond to special representations of $\IZ_n$
labeled by $p_\alpha$.

%%%%%%%%%%%%%%%%%%%%%%%%%%%%%%%%%%%%%%%%%%%%%%%%%%%%%%%%%%%%%%%%%%%%%%
%%%%%%%%%%%%%%%%%%%%%%%%%%%%%%%%%%%%%%%%%%%%%%%%%%%%%%%%%%%%%%%%%%%%%%

\newsec{Intersection form for fractional branes of the $\IC^d/\Gamma$ orbifold}

Following \refs{\DouglasHQ,\BerkoozIS,\BerkoozKM} we define the brane intersection form
on boundary conditions $a,b$ to be: 
\eqn\defind{I_{ab}=\tr_{R,ab} (-1)^F q^{L_0-{c\over 24}}   .}
Here the trace is over the states of the open string suspended
between D-branes which correspond to representations of $\Gamma$
labeled by $a$ and $b$ and $F$ is the worldsheet fermion
number.
The matrix $I_{ab}$ is actually an index, i.e. does not depend on
the modulus of the worldsheet cylinder $q$, and counts the difference
of positive and negative chirality Ramond ground states.

The computation of $I_{ab}$ for the spacetime supersymmetric
$\IC^d/\IZ_n$ orbifolds can be found in \refs\bcr. 
We will describe here the computation 
along the same lines for $\IC^2/\IZ_{n(p)}$. We then give a simple 
general formula for $\IC^d/\Gamma$. 
The cylinder amplitude that is invariant under the orbifold
projection is
\eqn\ca{ I_{ab}={1 \over n} \sum_{g\in \IZ_n} \tr_a (g) \tr_b (g) \Tr_R
                  \left(g (-1)^F \,q^{L_0-{c\over 24}}  \right)               }
where $\tr_a (g)$ stands for the trace over the Chan-Paton factors;
it is convenient to choose a basis where the action of $g$ on Chan-Paton
factors is diagonal \DouglasSW.
Another ingredient,
\eqn\tttrrr{ \Tr_R \left(g \,q^{L_0-{c\over 24}}  \right)
          = \prod_{i=1,2} \Tr_{Z^{(i)}} \left(g \,q^{L_0-{c\over 24}}  \right)
                          \Tr_{\psi^{(i)},R} \left(g \,(-1)^F\,q^{L_0-{c\over 24}}  \right)   }
can be evaluated in the NSR formalism separately for worldsheet bosons and fermions.
The generator of the orbifold group $w$ acts on the complex boson as
\eqn\orbactx{w:\;Z^{(i)} \ra e^{2\pi i \nu_i} Z^{(i)}     ,}
where $\nu_1=1/n,\; \nu_2=p/n$.
Hence,
\eqn\trz{ \Tr_{Z^{(i)}} \left(g \,q^{L_0-{c\over 24}}  \right)=
          q^{-{1\over 12}}\prod_{n=1}^\infty\, (1-e^{2\pi i\nu_i}q^n)^{-1}(1-e^{-2\pi i\nu_i}q^n)^{-1} }
The fermion part can be evaluated in the similar manner.
The nontrivial ingredient is now the presence of zero modes
in the R sector.
These zero modes satisfy anticommutation relations:
\eqn\acm{ \{\psi_0^{(i)}, \bar\psi_0^{(j)}\}=\delta^{ij},\qquad  
 \{\psi_0^{(i)}, \psi_0^{(j)}\}=\{\bar\psi_0^{(i)},
          \bar\psi_0^{(j)}\}=0       .}
The ground state must furnish the representation of this algebra;
it is a spacetime spinor labeled by the eigenvalues of
$S^{(i)}={\bar\psi_0^{(i)}} \psi_0^{(i)}-{1\over 2}$
which take values $s^{(i)}=\pm 1/2$.
Then  $g=w^s\in\IZ_{n(p)}$ acts on the ground state as
\eqn\orbactgs{ g=w^s:\; |s^{(1)},s^{(2)}\rangle \ra
   \exp\left({\sum_{i=1,2} 2 \pi i s^i \nu^i s}\right)   |s^{(1)},s^{(2)}\rangle   }
Note that $p$ must be odd in order for the orbifold group to
be $\IZ_n$, $w^n=1$.
The fermion part of the trace is
\eqn\trpsi{ \Tr_{\psi^{(i)}} \left(g (-1)^F\,q^{L_0-{c\over 24}}  \right)= 2\sin\left(\pi\nu_i\alpha\right)
     q^{1\over 12}\prod_{n=1}^\infty\, (1-e^{2\pi i\nu_i}q^n)(1-e^{-2\pi i\nu_i}q^n) }
Combining everything together, the intersection form is
\eqn\om{I_{ab}={4 \over n}\sum_{s=0}^{n-1}\exp\left({2\pi i (a-b)s \over n}\right)
                           \sin\left({\pi s\over n}\right)\sin\left({\pi s p\over n}\right) }
Here the value of $g=w^s$ in the representation of $\IZ_{n(p)}$ labeled
by $a$ is $e^{2\pi i a s\over n}$.
One observes that $I_{ab}$ is not invariant under $p\ra p\pm n$.
This is not really surprising, as type II theory is not invariant
under such a shift.
\foot{The situation with type 0 is more subtle.
The closed string sector of type 0 theory is invariant under $p\ra p\pm n$;
however the orbifold action in the open string sector does change, hence
the theory is again defined by specifying $p\;{\rm  mod}\; 2n$.}
The intersection form \om\ can be easily evaluated; the result is
\eqn\oma{I_{ab}=\delta_{a-b-{1-p\over 2}}+\delta_{a-b+{1-p\over 2}}
           -\delta_{a-b-{1+p\over 2}}-\delta_{a-b+{1+p\over 2}}         }
where
\eqn\defdm{\delta_a\equiv\delta_{a, 0\;\mn}.}
Note that the arguments of delta functions in \oma\ are
always integers, thanks to the requirement that $p$ is odd.

The above quadratic form has a natural interpretation in K-theory,
which easily generalizes to all values of $d$ and finite groups 
$\Gamma$ acting linearly on $\IC^d$. Let $R(\Gamma)$ denote the representation ring
of $\Gamma$. As we have seen, it is necessary to choose a lift of 
$\Gamma$ to the spin group, so we assume $\Gamma$ is a discrete subgroup 
of the spin group.  Since  $\Gamma$ acts on the spin bundle  
we can define the equivariant  index, and hence   a bilinear pairing on $K_\Gamma^0(X)$:
\eqn\indxpr{
(E,F) \to  Ind_\Gamma(\Dsl_{E^*\otimes F}) \in R(\Gamma)
}
Composing with the projection  $\pi_*: R(\Gamma) \to \IZ$ defined by
the projection to the trivial representation we obtain a
bilinear pairing on  $K_\Gamma^0(X)$ with values in $\IZ$. This is the natural
pairing on branes defined by elements of equivariant
$K$-theory. Now let us specialize to $X = \IC^d$.
Quite generally, $K_\Gamma(X) \cong K_\Gamma(pt) \cong R(\Gamma)$ whenever
$X$ is equivariantly contractible to a point \segal.
Evaluating the topological index we obtain the pairing on
fractional branes  thought of as elements of
$K_\Gamma(\IC^d) \cong R(\Gamma)$. This pairing is given by
\eqn\kpair{
(\rho_1, \rho_2)  = \pi_*\biggl( \bar \rho_1\otimes \rho_2 \otimes (S^+ - S^-) \biggr)
}
Here $S^\pm$ are the chiral spin representations on $\IC^d$ regarded
as $\Gamma$-modules.  This pairing is precisely that computed above using
D-brane techniques.

%%%%%%%%%%%%%%%%%%%%%%%%%%%%%%%%%%%%%%%%%%%%%%%%%%%%%%%%%%%%%%%%%%%%%%%%%%%%%%%%%%%%%%%%%%%%
%%%%%%%%%%%%%%%%%%%%%%%%%%%%%%%%%%%%%%%%%%%%%%%%%%%%%%%%%%%%%%%%%%%%%%%%%%%%%%%%%%%%%%%%%%%%

\newsec{Finding the change of basis}

\subsec{Statement of the main result}

The intersection matrix $I_{ab}$ should be interpreted
as the intersection form evaluated on the elements of the equivariant
K-theory of the orbifold.
We denote this   by
\eqn\ixtion{I_{ab}=(e_a,e_b)   }
We will demonstrate the existence of an {\it integral} invertible linear transformation
\eqn\etof{f_a=U_{a}^{~b} e_b}
which block-diagonalizes $I$
\eqn\ibd{
  F=-UIU^T =\pmatrix{
~0& ~0& ~0\cr
~0& ~C& ~0\cr
~0& ~0& ~C'}
}
In \ibd\ the first row and the first column of zeroes correspond
to the D0 brane, which is free to move off the orbifold fixed point.
The next $r$ rows and columns correspond to matrix elements between the
special representations, $e_{p_r=1},e_{p_{r-1}},\ldots,e_{p_1}$.
Moreover, $C$ is the generalized $r\times r$ Cartan matrix \hjint\ which describes the
geometric intersection of the exceptional divisors in the
minimal \HJ\ resolution.
We will find that  $U$ has the simple   form
\eqn\uut{
  U =\pmatrix{
~1& ~1_{1\times r} & ~1_{1\times r'} \cr
~0_{r\times 1} & ~1_{r\times r} & ~u\cr
~0_{r'\times 1} & ~0_{r' \times r} & ~1_{r'\times r'} }
}
where $u$ is an $r \times r'$ matrix of integers.

%%%%%%%%%%%%%%%%%%%%%%%%%%%%%%%%%%%%%%%%%%%%%%%%%%%%%%%%%%%%%%%%%%%%%%%%%%5
\subsec{Examples}
Here we provide some examples which illustrate the general claim
outlined above.
The first example is rather trivial, $n(p)=n(-1)$ which gives rise
to the supersymmetric orbifold.
The continued fraction is $[2^{n-1}]$, so all generators survive the
GSO projection, as expected.
As $r=n-1$, all  $e_i$ correspond to special representations, and
no additional change of basis is required.

The next example is $n(p)=n(1-n)$.
(Note that $n$ must be even, so that $p=1-n$ is odd.)
The action on worldsheet bosons here is the same as in the $n(1)$
orbifold, but the latter suffers from the merciless GSO projection 
which eliminates the generator of the chiral ring.
The continued fraction is $[n]$.
It is not difficult to show that
\eqn\uexb{ u=\left(2,3,\ldots,{n\over 2},-{n\over 2}+1,-{n\over 2}+2,\ldots,-1\right)   }
is the solution with all required properties.

Our next example is $n(p)=(3m+1)(-3)$.
The continued fraction is $[2^{m-1},4]$.
The solution is given by
\eqn\uexc{
  u =\pmatrix{
~2&  ~-1& ~1& ~-1& ~1& ~-1& ~\ldots \cr
~0&  ~1& ~1&  ~0& ~0&  ~0& ~\ldots \cr
~0&  ~0& ~0&  ~1& ~1&  ~0& ~\ldots \cr
~\ldots}
}
%%%%%%%%%%%%%%%%%%%%%%%%%%%%%%%%%%%%%%%%%%%%%%%%%%%%%%%%%%%%%%
\subsec{The proof of the main result}
 
Let us omit the first row and the first column, which corresponds
to restricting to the subspace spanned by the fractional branes,
$e_i,\,i=1\ldots n-1$.
(The first row and column contain the overlaps of
%%%%
$e_i$ with  $D0-\sum_{i=1}^{n-1} e_i$,
where $D0$ stands for the D0 brane which can move off the orbifold singularity.)
In the following the overlap matrices will be always restricted
to this subspace, and will be denoted by the same letters as  
their unrestricted counterparts above.
We will prove that for the restricted matrix $I$, written in the
basis
\eqn\ebasis{\{e_{p_1},\ldots,e_{p_r},\{e_{\nu}\}  \}     }
where $\nu$ labels Coulomb representations, i.e.
runs over $2,\ldots n-1$ with $p_\alpha$ excluded,
there exists a matrix $U$ of the form
\eqn\uuta{
  U =\pmatrix{
~1& ~u& \cr
~0& ~1 }
}
with $u$ being an $r\times(n-r-1)$ matrix of integers, such that
\eqn\ibda{
  -UIU^T =\pmatrix{
~C&   ~0\cr
~0& ~C'}
}
where $C$ is the generalized Cartan matrix.
We will actually prove the inverse of \ibda, which is equivalent,
since all the matrices there are non-singular.
Denote
-\eqn\iidef{-I^{-1}=\pmatrix{
~\tc&   ~\tx\cr
~\tx^T&   ~\tc'}
}
Using
\eqn\uinv{
  U^{-1} =\pmatrix{
~1& ~-u& \cr
~0& ~1 }
}
and a similar formula for the transpose, we can write
$-(U^T)^{-1} I^{-1} U^{-1}$ as
\eqn\ibdb{
-(U^T)^{-1} I^{-1} U^{-1}=
\pmatrix{
~\tc&   ~-\tc u +\tx \cr
~-u^T \tc+\tx^T& ~-u^T(-\tc u+\tx)-\tx^T u+\tc'}
}
To prove \ibda\ (or rather its inverse)
it is sufficient to prove two statements.
First, $\tc=C^{-1}$, or
\eqn\igc{
(-I^{-1})_{p_\alpha p_\beta}=(C^{-1})_{\alpha\beta}
	 = \cases{ {1\over n} q_\alpha p_\beta &
		$1\leq \alpha\leq \beta\leq r$ \cr
		& \cr
	{1\over n} p_\alpha q_\beta &
		$1\leq \beta \leq \alpha\leq r$ }\quad .
}
where we used an expression for $C^{-1}$ derived in \MartinecWG.
Second, 
\eqn\uctx{u=C \tilde x  }
is a matrix of integers.
Let us start with the first statement.

We want to compute $(-I^{-1})_{jk}$.
Define
\eqn\phidef{\phi_{jk}=
   {\exp\left({2\pi i j k\over n}\right)\over \sqrt{n}}=
       {w^{jk}\over \sqrt{n}}, \qquad j,k=1,\ldots,n-1.}
and
\eqn\sigmabigdef{\Sigma={\rm diag}\{\sigma_j\}, \qquad
           \sigma_j=4 \sin\left({\pi j \over n}\right) \sin\left({\pi j p\over n}\right) .}
In matrix notation the inverse of \om\ reads
\eqn\oma{I^{-1}=\phi^{-1} \Sigma^{-1} ({\bar\phi})^{-1},}
where bar denotes complex conjugation.
The matrix $\phi$ is not hard to invert; the result is
\eqn\phiinverse{(\phi^{-1})_{jk}={\bar\phi}_{jk}-{1\over \sqrt{n}} }
This can be verified by making use of the identities
\eqn\phiphi{\phi_{jk}{\bar\phi}_{kl}=\delta_{jl}-{1 \over n} .}
and
\eqn\phisum{\sum_{k=1}^{n-1}\phi_{jk}=-{1 \over \sqrt{n}} .}
Combining \phiinverse\ and \oma, we have
\eqn\iformula{(-I^{-1})_{jk}=N_{jk}(n;p_1)}
where we define
\eqn\njkdef{ N_{jk}(n;p_1)=
{1\over n}\sum_{s=1}^{n-1} {(e^{-2\pi i j s\over n}-1)
  (e^{2\pi i k s\over n}-1)\over 4 \sin\left({\pi s\over n}\right)
                                    \sin\left({\pi (n-p_1)s\over n}\right)}   }
To deal with this formula we draw inspiration from
appendix A of \hkmm.
Consider the meromorphic function
\eqn\fdef{f_{jk}(z)=-{ (e^{-2\pi i j z}-1) (e^{2\pi i k z}-1)\over
             4 \sin(\pi z)\sin(\pi p_1 z)\sin(\pi n z)}            }
We consider contour integrals of this function around
the contour given by $\CC= C_1 - C_2 - C_3 + C_4$ where
$C_1$ runs along $1-\epsilon + iy $, $-\Lambda\leq y \leq \Lambda$,
$C_2$ runs along $x + i \Lambda$, $-\epsilon\leq x \leq 1-\epsilon$,
$C_3$ runs along $-\epsilon + iy $, $-\Lambda\leq y \leq \Lambda$,
and
$C_4$ runs along $x - i \Lambda$, $-\epsilon\leq x \leq 1-\epsilon$.
Here $\epsilon<1/n$.

Note that 
by contour deformation, and evaluation of residues we learn that 
\eqn\finta{
{1\over 2i} \oint_{\CC}  f_{jk}(z)dz =  -{jk\over p_1 n} +N_{jk}(n;p_1)+N_{jk}(p_1;n)  
 .}
On the other hand, we can explicitly evaluate the integrals along the countour 
$C_1-C_2 -C_3+C_4$ as follows. 
Note that $f_{jk}(z+1)=f(z)$ as long as $p=p_1-n$ is an odd number,
which is precisely the condition for type II theory to be well defined.
Hence, the integrals over $C_1$ and $-C_3$ cancel each other.
The integrals along $C_2, C_4$ are not zero in general. However, 
along $C_4$, we may expand $f_{jk}$ in inverse powers of $\xi:=e^{i \pi z} $, 
which has a large absolute magnitude. The result is that the contribution 
of the segment $C_4$ is 
\eqn\intcf{
{1\over 2i} \int_{-\epsilon}^{1-\epsilon} dx \xi^{2k-1-p_1-n} 
{ (1- \xi^{-2j}) (1-\xi^{-2k} )\over (1-\xi^{-2})(1-\xi^{-2p_1}) 
(1-\xi^{-2n}) } 
}
A similar formula holds for $C_2$. 
Note that for $p$ odd, the expansion is in powers of $\xi^{-2}$ and hence 
the integral vanishes unless there is a term $\sim \xi^0$. We therefore 
define some coefficients 
\eqn\defas{
{ (1- \xi^{-2j}) (1-\xi^{-2k} )\over (1-\xi^{-2})(1-\xi^{-2p_1}) 
(1-\xi^{-2n}) } := \sum_{\ell=0}^\infty A_\ell \xi^{-\ell} 
}
and define moreover $A_\ell=0$ when $\ell<0$. Note that these coefficients 
are {\it integers}. In terms of these integers we may 
write
\eqn\fint{
{1\over 2i} \oint_{C_1-C_2 -C_3+C_4} dz f_{jk}(z) = - A_{(2j-1-p_1-n)} - A_{(2k-1-p_1-n)}
}
Note, especially, that the integral \fint\ vanishes if $j,k< {p_1+n+1\over 2}$. 

Comparing \finta\ to \fint, we obtain a recursion formula for $N_{jk}(n;p_1)$. 
We now combine this recursion formula with the observation that 
\eqn\nrel{   N_{jk}(p_1;n)=-N_{jk}(p_1;p_2)         }
where $p_2$ is defined to be $p_2=a_1p_1-n$, and 
which holds when $a_1$ is an even integer.
As long as $j,k$ satisfy the relevant bound so that \fint\ vanishes 
 we can now write  
\eqn\njkstepa{ N_{jk}(n;p_1)= - N_{jk}(p_1;n)+{jk\over p_1 n}=
                    N_{jk}(p_1;p_2)+{jk\over p_2 p_1}+{jk\over p_1 n}=\ldots}
where we keep getting $N_{jk}(p_\alpha;p_{\alpha+1})$ with higher and higher $\alpha$'s.
For this construction to work, all $a_\alpha$'s must be even integers.
Recall that this is precisely the condition for the chiral ring generators to survive
the GSO projection.
%In addition to the terms of the form ${jk\over p_\alpha p_{\alpha-1}}$, \njkstepa  might
%contain integer terms coming from \fint.
If either $j$ or $k$ is equal to $p_\alpha$, the process terminates
when we reach $N_{jk}(p_\alpha;p_{\alpha+1})=0$.

Consider $\tc_{\alpha\beta}=I^{-1}_{j=p_\alpha,k= p_\beta}$.
Suppose $\alpha\le \beta$ which implies $j=p_\alpha\ge k=p_\beta$.
The recursion process in \njkstepa\ terminates at $N_{jk}(p_\alpha;p_{\alpha+1})=0$.
There are no additional integer contributions since
$j,k \leq p_1 <{n+p_1+1\over 2}$.
That is, we have
\eqn\tcab{ \tc_{\alpha\beta} = p_\beta p_\alpha \sum_{\gamma=1}^\alpha
                      {1\over p_\gamma p_{\gamma-1}}  .}
The desired equality, \igc, follows from
\eqn\sumpp{ p_\alpha \sum_{\gamma=1}^\alpha {1\over p_\gamma p_{\gamma-1}} ={q_\alpha \over n}    .}
This can be proven by induction.
Define
\eqn\bigxdef{ H_\alpha=p_\alpha \sum_{\gamma=1}^\alpha {1\over p_\gamma p_{\gamma-1}}   }
and note that $H_1=1/n=q_1/n$, $H_2=a_1/n=q_2/n$.
Then it is not hard to prove that the induction hypothesis
\eqn\xproofa{ H_\alpha=a_{\alpha-1} H_{\alpha-1}-H_{\alpha-2}        }
implies
\eqn\xproofb{ H_{\alpha+1}=a_{\alpha} H_{\alpha}-H_{\alpha-1}       . }
To summarize,
\eqn\icpf{  \tc_{\alpha\beta}={q_\alpha p_\beta \over n}, \qquad \alpha\le\beta   .}
When $\alpha\ge\beta$, the sequence in \njkstepa\ terminates at $N_{jk}(p_\beta;p_{\beta+1})$
and so $\alpha$ and $\beta$ in \icpf\ are interchanged.
This concludes the proof of \igc.

To prove that $u=C \tx$ is a matrix of integers, write this as
\eqn\ucx{  u_{\alpha \nu}=C_{\alpha\beta} \tx_{\beta \nu}=
      -\tx_{\alpha-1, \nu}+a_\alpha \tx_{\alpha, \nu}-\tx_{\alpha+1,\nu}
        }
where index $\nu$ labels Coulomb representations, i.e. runs
over $2,\ldots,n-1$, with $p_1,\ldots,p_r$ excluded. 
The computation of  $\tx_{\alpha, \nu}=N_{p_\alpha,\nu}(n;p_1)$ proceeds via reduction,
just as in the
case of $\tc_{\alpha\beta}$.
The process in \njkstepa\ has additional integer 
contributions from the coefficients $A_\ell$ in \fint\ but still 
 terminates at $N_{p_\alpha,\nu}(p_\alpha;p_{\alpha+1})=0$.
However it is no longer true that $\nu<{p_\alpha+p_{\alpha+1} \over 2}$, and hence 
some of the coefficients $A_\ell$ will be nonzero. Nevertheless, we can say that 
\eqn\tcfla{  \tx_{\beta, \nu} = {q_\beta \cdot \nu\over n} + b_{\beta,\nu}, }
where $b_{\beta,\nu}$ is a sum of coefficients of the type $A_\ell$. In particular, 
they are integers. Substituting this into \ucx\ gives
\eqn\ufinal{  u_{\alpha \nu}= -b_{\alpha-1,\nu}+a_\alpha b_{\alpha,\nu}-b_{\alpha+1,\nu} \in \IZ   }
This completes the proof.

%%%%%%%%%%%%%%%%%%%%%%%%%%%%%%%%%%%%%%%%%%%%%%%%%%%%%%%%%%%%%%%%%%%%%%%%%%%%%%%%%%%%%%%%%%%
%%%%%%%%%%%%%%%%%%%%%%%%%%%%%%%%%%%%%%%%%%%%%%%%%%%%%%%%%%%%%%%%%%%%%%%%%%%%%%%%%%%%%%%%%

\newsec{Discussion of some loose ends  }

\subsec{Coupling to the bulk graviton}
The mass of a fractional brane can be measured by the overlap
of the corresponding boundary state with the bulk graviton and
is given by 
\eqn\mfr{m_{1/ n}={m_0 \over n}  ,}
where $m_0=1/g_s l_s$ is the mass of a D0 brane.
Higgs branch branes which wrap minimal cycles in the resolved
geometry come from the linear combination of fractional branes
determined by \etof.
As explained in the previous section, to each Higgs representation of
the orbifold group $\IZ_n$, labeled by $p_\alpha$, we can associate a linear combination
of fractional branes $e_i$,
\eqn\fb{  h_{ \alpha} = e_{p_\alpha} + \sum_\nu u_{\alpha}^{~\nu}e_\nu   }
According to \mfr, the mass of the corresponding boundary state is
given by
\eqn\massf{  {m_\alpha\over m_{1/n}}= 1+ \sum_\nu  u_{\alpha}^{~\nu}   .}
The computation of the sum in \massf\ is similar in spirit to
the computations performed in the previous section.
We start by noting 
\eqn\sumi{ 
  \sum_{i=1}^n \sum_{\beta=1}^r C_{\alpha\beta}(-I^{-1})_{p_\beta i}=
              \sum_{\gamma=1}^r \sum_{\beta=1}^r C_{\alpha\beta} (C^{-1})_{\beta\gamma} + 
               \sum_\nu u_{\alpha}^{~\nu}       ,}
where we used \iidef, \igc\ and \uctx.
Substituting \sumi\ into \massf, we obtain
\eqn\massfa{  {m_\alpha\over m_{1/n}}=  \sum_\beta 
               C_{\alpha\beta} \sum_{i=1}^n (-I^{-1})_{p_\beta i}  }
where we used
$\sum_{\beta=1}^r C_{\alpha\beta} (C^{-1})_{\beta\gamma}=\delta_{\alpha\gamma}$.
To compute $\sum_{i=1}^n (-I^{-1})_{p_\beta i}$ we perform summation over $i$
in the expression for $I^{-1}$ [equations \iformula\ and \njkdef]; 
the result is
\eqn\iim{ \sum_{i=1}^n (-I^{-1})_{p_\beta i}= n M_\beta(n;p) }
where we define
\eqn\mbetadef{ M_\beta(n;p_1)=
 -{1\over n} \sum_{s=1}^{n-1} {(e^{-2\pi i p_\beta s\over n}-1)
  \over 4 \sin\left({\pi s\over n}\right)
                                    \sin\left({\pi (n-p_1)s\over n}\right)}   }
This quantity is computed similarly to $N_{jk}(n;p_1)$-
we again consider integral along $\CC$ of the meromorphic function
\eqn\hdef{h_\beta(z)={ (e^{-2\pi i p_\beta z}-1) \over
             4 \sin(\pi z)\sin(\pi p_1 z)\sin(\pi n z)}            }
The analog of \finta\ is now
\eqn\hinta{
                     0 =  -{p_\beta^2\over 2 p_1 n} +M_\beta(n;p_1)+M_\beta(p_1;n)  
 .}
where the first term in the right-hand side comes
from the residue at $z=0$, and the total integral is
zero because $2 p_\beta< n+p_1+1$.
Next we can perform the recursion process, thanks to
the identity $M_\beta(n;p_1)=-M_\beta(p_1;p_2)$.
The recursion terminates at $p_\beta$; combining everything together we have
\eqn\mfinal{  {m_\alpha\over m_{1/n}}= \sum_{\beta=1}^r {C_{\alpha\beta} p_\beta q_\beta\over 2}   }
This formula evaluates to positive integers for a variety
of $(n,p)$; however we have not bothered to prove this in general.

\subsec{Other values of $(n,p)$ and nonminimal resolutions}

The proof of the previous section makes heavy use of the 
hypothesis that $a_{\alpha}$ are all even, and this is the 
origin of our restriction on the values of $n,p$ mentioned in 
the introduction. 
What happens when some of the partial quotients $a_\alpha$ are odd? 
In this case   some of the generators are projected out in the type II 
string and the above story becomes more complicated.
In   string theory language, the geometry cannot be resolved
by turning on localized tachyons. We expect that one can perform a 
partial block diagonalization involving 
fractional branes at remaining singularities and Higgs branch branes 
associated to smooth cycles. But we have not carried this out.

Similar remarks apply to nonminimal resolutions. In this case one of the 
partial quotients will be equal to one, which is odd. For example, 
if we   blow up a cycle that corresponds to
a generator
\eqn\wprod{ W=W_j W_{j+1} ,}
then, as shown in \hkmm\MartinecWG, the continued fraction 
$n/p_1 = [a_1,\ldots,a_r]$ is replaced by 
\eqn\newcont{
{n\over p_1} = [a_1,\ldots,a_j+1,1,a_{j+1}+1,\ldots,a_r]. 
}
Hence, we see that only a subset of orbifold resolutions
can be realized within type II theory.
This is again a  consequence of the chiral GSO projection.

\subsec{Type 0 strings}

It is natural to ask what happens in type 0 string
theory where the diagonal GSO acts trivially on the
chiral ring.
Consider the fractional brane overlap matrix in the type 0 case. 
The number of D-branes is now doubled.
However the boundary states of type 0 are related to the
boundary states of type II in a simple way.
Namely, the so-called ``electric'' ($\upsilon$) and ``magnetic'' ($o$) states
of type 0 have the form \bcr
\eqn\embs{
\eqalign{
&|a,\upsilon\rangle = \sum_s B_a^s
      {1 \over \sqrt{2} } \left( |s,NS,+\rangle\rangle + |s,R,+\rangle\rangle\right);\cr
&|a,o\rangle = \sum_s B_a^s
      {1 \over \sqrt{2} } \left( -|s,NS,-\rangle\rangle + |s,R,-\rangle\rangle\right)
}}
where $|s,NS,\pm\rangle\rangle$, $|s,R,\pm\rangle\rangle$ are
Ishibashi states that couple to closed strings in the $s$
twisted sector and $B_a^s$ are   coefficients which can be
determined from the Cardy condition \bcr.
Type II boundary states are of the form
\eqn\twobs{
|a,II\rangle = \sum_s B_a^s
      {1 \over 2 } \left( |s,NS,+\rangle\rangle -|s,NS,-\rangle\rangle
          + |s,R,+\rangle\rangle+|s,R,-\rangle\rangle\right)
.}
Comparing \embs and \twobs, we infer that the type 0 intersection form
is given by
\eqn\izero{ I_0=\pmatrix{
~0& ~I\cr
~I& ~0 }
}
where $I$ is the type II intersection form. Physically, chiral fermions 
only exist when branes of opposite type intersect.

It is worthwhile remarking that the above doubling of branes, together
with the intersection form \izero\  is extremely natural from the K-theoretic
viewpoint. Type 0 is an orbifold of the type II string by 
$(-1)^{F_{st}}$, where $F_{st}$ is spacetime fermion number \DixonIZ. 
If a group $G$ acts trivially on a space $X$ then there is a natural
isomorphism $K_G(X) \cong K(X)\otimes R(G)$ given by
\eqn\segalhom{
E \to \oplus_a  {\rm Hom}( E_a,E) \otimes \rho_a
}
where $E_a = X \times \rho_a$ is a trivial $G$ bundle   associated to the
irrep $\rho_a$ of $G$.
If we regard the type 0 string as an orbifold of the type II string by
$\IZ_2$ with generator $\sigma$ acting trivially on $X$ then we see that
$K_{\IZ_2}(X)  \cong K(X) \oplus K(X)\otimes \eta$
where $\eta$ is the nontrivial irrep of $\IZ_2$.
The lift of $\sigma$ to the spin group acts as $-1$; this is what we mean
by an orbifold by $(-1)^{F_{st}}$. Thus, the $K$-theory orientation class is
 $(S^+ - S^-) \otimes \eta$. This has an effect on the
 bilinear pairing.  Denoting by $(\cdot, \cdot )$
the tensor product of  the natural pairings on  $K(X) \otimes R(\IZ_2) $
the index pairing on $K_{\IZ_2}(X) $ is 
\eqn\typzpr{
Ind(E_1, E_2) = (E_1, E_2 \otimes \eta )
}
for $E_1,E_2 \in K(X)\otimes R(\IZ_2)$.
This is precisely the intersection matrix  \izero.

In the type 0 case the intersection form \izero\ is composed out of
the type II intersection matrices.
One peculiar feature of this theory is that the closed
string sector is insensitive to the shift $p \ra p+n$.
Hence, we expect the same \HJ\ space as a resolution of the
%%%%
orbifolds $\IC^2/\IZ_{n(p)}$ and $\IC^2/\IZ_{n(p+n)}$.
However the intersection form $I$ is certainly different in these
two cases.
Consider, for example,  the supersymmetric orbifold with $p=-1$, where
the fractional branes simply evolve into  the
branes wrapping cycles in the resolved ALE space.
No additional change of basis is required in this case,
as the intersection form $I_0(p=-1)$ has the form \izero\ with $I$ being
the Cartan  matrix of the supersymmetric orbifold.
But the intersection form $I_0(p=n-1)$ is very different.
In the simple example $n=3$ it is possible to bring it
to the supersymmetric form by a change of basis, $I_0(p=-1)=U I_0(p=n-1) U^T$,
with $U$ being an invertible matrix of integers.
However it is rather difficult to construct such a change
of basis in general.
This should be contrasted with the type II case, where
the change of basis has a relatively simple form and the fractional
branes that correspond to Higgs representations naturally
become the branes wrapping cycles in the resolved geometry.

\subsec{Intersection matrix for Coulomb branch branes} 

So far, we have focused on the fractional branes and the Higgs branch branes. 
It is natural to ask what can be said about the Coulomb branch branes. In particular, 
given the change of basis \chgbasis\ we expect the pairing on Coulomb branch 
branes to be given by the matrix $C'$ in \ibd. 

In general, the intersection form for branes in Landau-Ginzburg models is 
difficult to compute, although some nontrivial progress has been made 
\refs{\HoriCK,\KapustinRC,\KapustinGA,\KapustinKT}. For this reason, 
we restrict our discussion to the case of $d=1$, i.e. $\IC/\IZ_n$ orbifolds. 
In this case, there is one Higgs branch brane, which may be identified with 
the $D0$ brane on the resolution of the space $\IC/\IZ_n$, and $(n-1)$ 
Coulomb branch branes described by a Landau-Ginzburg model with 
superpotential $W(X)$ given by an order $n$ polynomial \hkmm. 

When the superpotential is $W(X) = X^n$ we can think of the branes
in terms of those of the $\CN=2$ minimal conformal field theory.
The latter have a simple geometrical description described in
\MaldacenaKY, which, moreover, is nicely compatible with the
intersection form on branes. According to \MaldacenaKY,
the   branes may be pictured as oriented straight lines
in the unit disk (= parafermion target space)
  joining special points on the boundary of the disk.
These special points are at  integral
multiples of the angle $\phi_0 = {\pi \over n}$.  In the type 0 theory
the branes separate into two types - those
joining even multiples of $\phi_0$ and those joining
odd multiples of $\phi_0$. In the language of \MaldacenaKY,
these are the Cardy states $\vert \hat j, \hat n, \hat s\rangle$
with $\hat s = \pm 1$ and $\hat s=0,2$, respectively, and they
may be identified with the so-called ``electric'' and ``magnetic''
branes of the type 0 theory. As discussed in \MaldacenaKY\
  a generating set  of branes is given by the shortest chords.
These are
$e_s^{-}$ joining  $(2s-2)\phi_0$ to $2s \phi_0$,
$s=1,\dots, n$ while $e_s^+$ joins
$(2s-1)\phi_0$ to $(2s+1) \phi_0$, $s=1,\dots, n$.
As explained in \MaldacenaKY, two concatenated chords with
the same orientation are classically unstable to form a shorter chord.
Therefore, the state $e_1^\pm + \cdots+ e_n^\pm $, which may be pictured
as a polygonal ring, is unstable to shrinking into the disk. We interpret
this brane as the Higgs branch brane of the spacetime, and the
relation $e_1^\pm + \cdots + e_n^\pm =0$ of \MaldacenaKY\ as the relation
defining the space of Coulomb branch branes.
The intersection form can be computed from the boundary state.
It turns out that   $(e_s^- , e_{s'}^-) = (e_s^+ , e_{s'}^+) =0$,
while the intersection form of odd-branes with even-branes is
simply the geometrical oriented intersection number. In particular,
$(e_s^- , e_{s'}^+)= \delta_{s,s'}$.
Thus the intersection form is simply
\eqn\doneint{
\pmatrix{0_{n\times n} & 1_{n\times n} \cr -1_{n\times n}  & 0_{n\times n} \cr}
}
Note that the change of basis $e_s^+ \to e_{s+1}^+$ on the odd 
branes brings the intersection matrix to the form 
\eqn\chgbsig{
\pmatrix{\CS & 0_{n\times n} \cr 0_{n\times n} & 1_{n\times n}\cr} \pmatrix{0_{n\times n} & 1_{n\times n} \cr -1_{n\times n} & 0_{n\times n} \cr} 
\pmatrix{\CS^{tr} & 0_{n\times n} \cr 0_{n\times n} & 1_{n\times n}\cr}
 =\pmatrix{0 & \CS \cr -\CS^{tr} & 0 \cr}
}
so we neatly reproduce the intersection form of the fractional branes of
type 0.
For the type II projection we note that spectral flow from the NS to the R
sector takes $e_s^- \to e_{s+1}^+$ so the type II branes
have a basis   $e_s = e_s^- + e_{s+1}^+$. The intersection
form is   $(e_s, e_{s'}) = - \delta_{s+1,s'} + \delta_{s,s'+1}$, in
perfect conformity to that of the fractional branes computed in section 4.

When the Landau-Ginzburg polynomial is deformed to $W = X^n + a_{n-1}
X^{n-1} + \cdots + a_0$
the critical points separate.  Let us adopt the heuristic picture
advocated in \hkmm.
Then, associated with each critical point is a ``universe'' with its own
Higgs branch
brane. If the critical point is not Morse, then it has a collection of
Coulomb branes
associated to it. Since we have the unit matrix in the $12$ block in
\doneint\
there is no problem splitting the branes into collections for each
critical point.
Thus, for   finite RG flow the intersection form
 \doneint\ is preserved. For infinite RG flow, the Witten index changes,
as expected.

\subsec{Other open problems}

The present note raises some further interesting questions
for the future.

One evident open problem is the generalization of our 
discussion to $d>2$. We expect that similar techniques 
will apply but many - possibly very nontrivial - 
 details remain to be worked out.

It would be quite  illuminating to  construct boundary 
states directly within the framework of the gauged 
linear sigma model and literally follow 
the RG evolution of the boundary state.  
It would also be nice to 
understand the relation of the quantum McKay correspondence 
to the discussion of ``missing D-brane charges'' 
found in \MinwallaHJ. 

Finally, understanding real time dynamics of tachyon
condensation  as opposed to studying the RG flow is an 
important subject in need of clarification. 
See \GregoryYB\HeadrickYU\ for recent work on this 
topic. The generalization to $d=2$ appears to be quite 
nontrivial. It would be good to understand the fate of 
D-branes in these time-dependent backgrounds.

%\refs{\HoriCK,\HoriIC} ) 

\bigskip  
\noindent{\bf Acknowledgements:}  
We would like to thank A. Adams,  E. Diaconescu, M. Douglas, B. Florea
and S. Terashima for useful discussions. We thank D. Kutasov and
E. Martinec for comments on the draft.
This work is supported in part by DOE grant DE-FG02-96ER40949.

\listrefs

\end